\documentclass[aps]{revtex4}
\usepackage{amssymb,epsfig,latexsym}
\usepackage{amsmath,amscd}
\usepackage{graphicx}
\usepackage[normalem]{ulem}
\usepackage[dvips]{color}

\begin{document}

\title{White dwarfs in de Rham-Gabadadze-Tolley like massive gravity}
\author{B. Eslam Panah$^{1,2,3}$\footnote{%
email address: beslampanah@shirazu.ac.ir}}
\author{H.L. Liu$^4$\footnote{%
email address: heleiliu@xju.edu.cn}}
\affiliation{$^1$ Physics Department and Biruni Observatory, College of Sciences, Shiraz
University, Shiraz 71454, Iran\\
$^{2}$ Research Institute for Astronomy and Astrophysics of Maragha (RIAAM),
P.O. Box 55134-441 Maragha, Iran\\
$^{3}$ ICRANet, Piazza della Repubblica 10, I-65122 Pescara, Italy\\
$^{4}$ School of Physical Science and Technology, Xinjiang University,
Urumqi, 830046, China}

\begin{abstract}
The existence of possible massive white dwarfs more than the
Chandrasekhar limit ($1.45M_{\odot }$, in which $M_{\odot }$ is mass of the
sun) is a challenging topic. In this regard and motivated by the important
effect of massive graviton on the structure of white dwarfs we study the
white dwarfs in Vegh's massive gravity which is known as one of theories of
de Rham, Gabadadze and Tolley (dRGT) like massive gravity. First, we
consider the modified Tolman-Oppenheimer-Volkoff equation in this theory of massive gravity and
solve it numerically by using the Chandrasekhar's equation of state. Our
results show that the maximum mass of white dwarfs in massive gravity can be
more than the Chandrasekhar limit ($M>1.45M_{\odot }$ ), and this result
imposes some constraints on parameters of massive gravity. Then, we
investigate the effects of various parameters on other properties of the
white dwarfs such as mass-radius relation, mass-central density relation,
Schwarzschild radius, average density and Kretschmann scalar. Next, we study
dynamical stability condition for super-Chandrasekhar white dwarfs and show
that these massive compact objects enjoy dynamical stability. Finally, in
order to have a better insight, we compare the super-Chandrasekhar white
dwarfs with the obtained massive neutron stars in dRGT like massive theory
of gravity.
\end{abstract}

\maketitle

\section{Introduction}

General relativity (GR) is a successful theory of gravity
contains gravitons as massless spin-$2$ particles. GR predicted
some phenomena such as the gravitational deflection of light
around gravitational sources such as the Sun, which was confirmed
by Arthur Eddington. Nowadays, its direct application in form of
gravitational lensing is one of the indispensable tools in
astrophysics and cosmology. Another powerful prediction of GR is
the presence of gravitational waves, which was detected by the
advanced LIGO/Virgo collaboration \cite{Abbott}. In spite of these
successes at the large scales, GR cannot explain why our universe
is undergoing an accelerated cosmic expansion. Therefore, GR
without the cosmological constant term needs to be modified. Among
these modified theories of gravity, massive gravity can explain
the late-time acceleration without considering dark energy
\cite{Koyama,Chamseddine,Amico,Akrami,Hinterbichler2012}. In order to build up a
massive theory with a massive spin-$2$ particle propagation, one
can add an interaction term to the Einstein-Hilbert action. In
addition, Chamseddine and Volkov found that the effect of graviton
mass is equivalent to introducing a matter source in the Einstein
equations which can consist of several different types of matter;
a cosmological term, quintessence and also non-relativistic cold
matter \cite{Chamseddine}. Massive gravity modifies gravitational
effects by weakening it at the large scale comparing to GR. This
allows the universe to accelerate, but its predictions at small
scales are the same as GR. On the other hand, massive gravity will
result into graviton having a mass of $m$ which in case of
$m\rightarrow 0$, the effect of massive gravity is vanished and
this theory reduces to GR. In addition, it was shown that the
graviton mass is very small in the usual weak gravity
environments, but becomes much larger in the strong gravity regime
such as; black holes and compact objects \cite{ZhangZ}.\
Accordingly, there were numerous developments in the massive
gravity theories in recent years \cite{Fierz,Van,van
Dam,Zakharov,Vainshtein,BD}. On the other hand, recent
observations by the advanced LIGO/Virgo collaboration has put a
tight bound on the graviton mass \cite{Abbott,AbbottII}, however
cannot rule out the possibility of nonzero mass. Also, there are
other theoretical and
empirical limits on the graviton's mass (see refs. \cite%
{massGI,massGII,massGIII,massGIV,massGV}, for more details). Thus
one may motivate to investigate the effects of considering the massive
gravitons on various branches related to gravitation.

Fierz and Pauli in 1939 introduced a class of massive gravity theory in flat
background \cite{Fierz}. In other words, Fierz and Pauli added the
interaction terms at the linearized level of GR, which this theory is known
as Fierz and Pauli massive (FP massive) gravity. Then van Dam, Veltman and
Zakharov found out that FP massive gravity suffers from discontinuity which
is known as van Dam-Veltman-Zakharov (vDVZ) discontinuity \cite{Van,van
Dam,Zakharov}. In order to remove the vDVZ discontinuity, Vainshtein
showed that such discontinuity appears as a consequence of working with the
linearized theory of GR, and then a mechanism for the nonlinear massive
gravity was introduced by him \cite{Vainshtein}. On the other hand,
Boulware and Deser explored that such nonlinear generalizations usually
generate an equation of motion which has a higher derivative term yielding a
ghost instability in the theory which is known as Boulware-Deser (BD) ghost
\cite{BD}. However, these problems, arising in the construction of the
massive gravity have been resolved in the last decade by introducing St\"{u}%
ckelberg fields \cite{Hamed}. This allows a class of potential energies
depending on the gravitational metric and an internal Minkowski metric
(reference metric). In refs. \cite{de RhamI,de RhamII}, de Rham, Gabadadze
and Tolley (dRGT) introduced a new version of massive gravity which is free
of vDVZ discontinuity and BD ghost in arbitrarily dimensions \cite{Hassan}.
Although the equations of motion have no higher derivative term in the dRGT
massive gravity, finding exact solutions in this theory of gravity is
difficult. However, black hole solutions in dRGT massive gravity have been
obtained by some authors in refs. \cite{Cai,Kodama,Ghosh,Zou}. In the
astrophysics context, Katsuragawa et al. evaluated the neutron stars in this
theory and showed that the massive gravity leads to small deviation from the
GR \cite{Katsuragawa}. Mass-radius ratio bounds for compact objects in this
gravity have been obtained in ref. \cite{Kareeso}. M. Yamazaki et al.
discussed the boundary conditions for the relativistic stars in this theory
of gravity \cite{Yamazaki}. From cosmological point of view, bounce and
cyclic cosmology \cite{CaiI}, cosmological behavior \cite{Leon}, and another
properties have been studied by some authors in refs. \cite%
{Hinterbichler,Fasiello,Bamba}. On the other hand, the constraints imposed
by the Type Ia Supernovae (SNe Ia), Gamma Ray Bursts (GRBs), Baryon Acoustic
Oscillations (BAOs), Cosmic Microwave Background Radiation (CMBR) on the
massive gravity have been investigated in refs. \cite{CardoneR,HeisenbergR}.
In ref. \cite{Panpanich}, Panpanich and Burikham evaluated the effects of
nonzero graviton mass on the rotation curves of the Milky Way, spiral
galaxies and Low Surface Brightness galaxies. Rotation curves of the most
galaxies can be fitted well by considering the graviton's mass in the range $%
m\sim 10^{-21}-10^{-30}eV$ (see ref. \cite{Panpanich}, for more details).
Aoki and Mukohyama studied graviton mass as a candidate for dark matter \cite%
{AokiII}. Indeed, they showed that if LIGO detects gravitational waves
generated by preheating after inflation then the massive graviton with the
mass of\ $\sim 0.01GeV$ is a candidate of the dark matter. Cosmological
perturbations in massive gravity have been studied by some authors in refs.
\cite{PerturI,PerturII,PerturIII,PerturIV,PerturV}, and they obtained some
constraints on parameters of this theory by considering observational
cosmological data.

It is notable that, modification in the introduced reference metric in dRGT
theory leads to the possibility of introduction of different classes of dRGT
like massive theories \cite{de Rham}. One of theories was proposed by Vegh
which has applications in gauge/gravity duality \cite{Vegh}. Indeed this
theory is similar to dRGT massive gravity with a difference that its
reference metric is a singular one. Graviton in this massive gravity may
behave like a lattice and exhibits a Drude peak \cite{Vegh}. It was shown
that for arbitrary singular metric, this theory of massive gravity is
ghost-free and stable \cite{Zhang}. Black hole solutions in this gravity
have been obtained in refs. \cite{Cai2015,HendiJHEP2015}. The existence of
van der Waals like behavior in extended phase space for the obtained black
holes has been studied in this massive gravity by some authors in refs. \cite%
{PV,PVI,PVII,PVIII}. It was pointed out that it is possible to have a heat
engine for non-spherical black holes in massive gravity \cite{Heat}. In
addition, magnetic solutions in this dRGT like massive gravity have been
addressed in refs. \cite{Mag,MagI}. From the perspective of astrophysical,
the modified TOV equation by considering this theory of massive gravity was
obtained in ref. \cite{TOV massive}, and it was shown that the maximum mass
of neutron stars can be more than three times of solar mass. The existence
of a remnant for black hole in this theory of massive gravity has been
evaluated by Eslam Panah and et al in ref. \cite{Eslam}, They showed that
this remnant may help to ameliorate the information paradox.

As we know, the massive graviton leads to the modification of
long-range gravitational force. Therefore, one may expect that the
graviton mass could be comparable to the cosmological constant,
which could illustrate the accelerated expansion of the Universe
without introducing the cosmological constant (see refs.
\cite{Amico,NewII,NewIII,NewIV,NewV}, for more details).
It is very interesting to apply the dRGT like theories of
gravity (in this work we consider Vegh's approach of massive
gravity) to astrophysical phenomena. It is notable that,
construction the general framework which quantifies the deviations
from the predictions of the GR in strong-gravity regime is very
difficult (see ref. \cite{PsaltisLRR}, for more details). In
addition, it is very important that if we could conclude that
cosmological and astrophysical applications are compatible with
observations in a specific theory of modified gravity. According
to the above reasons, it is necessary to study the compact objects
in the massive gravity, especially dRGT like massive gravity as
astrophysical test of this gravity in strong-gravity regime.

On the other hand, in recent years, some peculiar type SNe
Ia: e. g. SN 2006gz, SN 2007if, SN 2009dc, SN 2003fg, have been
observed \cite {Howell,Scalzo,Hicken,Yamanaka} with exceptionally
higher luminosities. It has been suggested that the progenitor
mass to explain such SNe Ia stand in the range $2.1-2.8M_{\odot
}$~\cite{Silverman,Taubenberger}, which exceeds
significantly the Chandrasekhar mass limit about $1.45M_{\odot }$. Some authors explained such over luminous SNe Ia by
proposing the existence of super strong uniform magnetic fields
~\cite{Das}, rotation white dwarfs~\cite{Boshkayev}, electrical
charge distribution white dwarfs \cite{Liu}, and modification to
GR in white dwarfs \cite{DasJCAP}, to generate some
super-Chandrasekhar white dwarfs. Briefly, for explanting these
massive white dwarfs, we can consider two approaches. The first
approach: we can improve equation of state by adding magnetic
field~\cite{Das}. The second approach is related to modified TOV
equations \cite{Boshkayev,Liu,DasJCAP,liu2018}. According to the mentioned
reasons for modification of GR, in this work, we are going to
investigate the influence of\ Vegh's massive gravity
\cite{Vegh} on the properties and stability of the white dwarfs.

The article is organized as follows: after an introduction
about Vegh's massive gravity, we will present the modified TOV in
this theory of massive gravity. In Sec. III, we reintroduce the
Chandrasekhar's equation of state as a suitable equation of state.
Then, by considering the mentioned modified TOV, we will study the
properties of white dwarfs. We will evaluate another quantities
such as Schwarzschild radius, average density, the Kretschmann
scalar and dynamical stability of these white dwarfs. Then we
compare the properties of massive neutron stars with
super-Chandrasekhar white dwarfs in this gravity. Some closing
remarks are given in the last section.

\section{Basic Equations}

The action of dRGT like massive gravity is given by \cite{Vegh}
\begin{equation}
\mathcal{I}=\frac{1}{2\kappa }\int d^{4}x\sqrt{-g}\left[ \mathcal{R}%
+m^{2}\sum_{i}^{4}c_{i}\mathcal{U}_{i}(g,f)\right] +I_{matter},
\label{Action}
\end{equation}%
where $R$ and $m$ are the Ricci scalar and the graviton
mass, respectively. $\kappa =$\textbf{\ }$\frac{8\pi G}{c^{4}}$, and also $f$ and $%
g$ are a fixed symmetric tensor and metric tensor, respectively. In the
above relation, $I_{matter}$ is related to the action of matter. In
addition, $c_{i}$'s are free parameters of this theory which are
arbitrary constants. Their values can be determined according to theoretical
or observational considerations \cite{Amico,Akrami,Berezhiani,conformal}.
Also, $\mathcal{U}_{i}$'s are symmetric polynomials of the eigenvalues of $%
4\times 4$ matrix $K_{\nu }^{\mu }=\sqrt{g^{\mu \alpha }f_{\alpha \nu }}$
(for $4$-dimensional spacetime) where they can be written in the following
forms
\begin{eqnarray}
\mathcal{U}_{1} &=&\left[ \mathcal{K}\right] ,\;\;\;\;\;\;\;\mathcal{U}_{2}=%
\left[ \mathcal{K}\right] ^{2}-\left[ \mathcal{K}^{2}\right] ,  \notag \\
\mathcal{U}_{3} &=&\left[ \mathcal{K}\right] ^{3}-3\left[ \mathcal{K}\right] %
\left[ \mathcal{K}^{2}\right] +2\left[ \mathcal{K}^{3}\right] ,  \notag \\
\mathcal{U}_{4} &=&\left[ \mathcal{K}\right] ^{4}-6\left[ \mathcal{K}^{2}%
\right] \left[ \mathcal{K}\right] ^{2}+8\left[ \mathcal{K}^{3}\right] \left[
\mathcal{K}\right] +3\left[ \mathcal{K}^{2}\right] ^{2}-6\left[ \mathcal{K}%
^{4}\right] .  \notag
\end{eqnarray}%
where the square root in $\mathcal{K}$ stands for matrix
square root, i.e. $\mathcal{K}_{~~~\nu }^{\mu }=\left( \sqrt{\mathcal{K}}%
\right) _{~~~\lambda }^{\mu }\left( \sqrt{\mathcal{K}}\right) _{~~~\upsilon
}^{\lambda }$, and the rectangular bracket denotes the trace $\left[
\mathcal{K}\right] =\mathcal{K}_{~~~\mu }^{\mu }$.

Considering a spherical symmetric space-time in $4$%
-dimensional as%
\begin{equation}
ds^{2}=g_{\mu \nu }dx^{\mu }dx^{\nu }=H(r)dt^{2}-\frac{dr^{2}}{S(r)}%
-r^{2}h_{ij}dx_{i}dx_{j},~~i,j=1,2.  \label{Metric}
\end{equation}%
where $H(r)$ and $S(r)$ are unknown metric
functions, and $h_{ij}dx_{i}dx_{j}=\left( d\theta ^{2}+\sin ^{2}\theta
d\varphi ^{2}\right) $. By variation of Eq. (\ref{Action}) with
respect to the metric tensor $g_{\mu }^{\nu }$, the equation of motion for
massive gravity can be written as%
\begin{equation}
G_{\mu }^{\nu }+m^{2}\chi _{\mu }^{\upsilon }=\frac{8\pi G}{c^{4}}T_{\mu
}^{\nu },  \label{field1}
\end{equation}%
where $G$ is the gravitational constant, and also, $G_{\mu }^{\nu }$ and $c$%
\ are the Einstein tensor and the speed of light in vacuum, respectively. $%
T_{\mu }^{\nu }$ denotes the energy-momentum tensor which comes from the
variation of $I_{matter}$ and $\chi _{\mu \nu }$ is the massive term with
the following explicit form
\begin{eqnarray}
\chi _{\mu \nu } &=&-\frac{c_{1}}{2}\left( \mathcal{U}_{1}g_{\mu \nu }-%
\mathcal{K}_{\mu \nu }\right) -\frac{c_{2}}{2}\left( \mathcal{U}_{2}g_{\mu
\nu }-2\mathcal{U}_{1}\mathcal{K}_{\mu \nu }+2\mathcal{K}_{\mu \nu
}^{2}\right)  \notag \\
&&  \notag \\
&&-\frac{c_{3}}{2}(\mathcal{U}_{3}g_{\mu \nu }-3\mathcal{U}_{2}\mathcal{K}%
_{\mu \nu }+6\mathcal{U}_{1}\mathcal{K}_{\mu \nu }^{2}-6\mathcal{K}_{\mu \nu
}^{3})  \notag \\
&&  \notag \\
&&-\frac{c_{4}}{2}\left( \mathcal{U}_{4}g_{\mu \nu }-4\mathcal{U}_{3}%
\mathcal{K}_{\mu \nu }+12\mathcal{U}_{2}\mathcal{K}_{\mu \nu }^{2}-24%
\mathcal{U}_{1}\mathcal{K}_{\mu \nu }^{3}+24\mathcal{K}_{\mu \nu
}^{4}\right) .
\end{eqnarray}

Considering the white dwarf as a perfect fluid with the following
energy-momentum tensor as %
\begin{equation}
T^{\mu \nu }=\left( c^{2}\rho +P\right) U^{\mu }U^{\nu }-Pg^{\mu \nu },
\label{EMTensorEN}
\end{equation}%
where $P$ and $\rho $ are the pressure and
density of the fluid which are measured by the local observer, respectively,
and $U^{\mu }$ is the fluid four-velocity. The nonzero components
of the energy-momentum tensor for perfect fluid are %
\begin{equation}
T_{0}^{0}=c^{2}\rho ,~~~T_{1}^{1}=T_{2}^{2}=T_{3}^{3}=-P.
\end{equation}

In order to obtain exact static spherical black hole solutions, the
appropriate ansatz for the reference metric was introduced in the form; $%
f_{\mu \nu }=diag(0,0,C^{2}h_{ij})$, see \cite{Cai2015,HendiJHEP2015}%
, for more details. In this work, we intend to obtain static spherical
solutions similar to static spherical black hole solutions of massive
gravity. So, we consider the mentioned appropriate ansatz for the reference
metric $f_{\mu \nu }$ in $4$-dimensional spacetime, which is given
as%
\begin{equation}
f_{\mu \nu }=diag(0,0,C^{2},C^{2}\sin ^{2}\theta ),  \label{referenceM}
\end{equation}%
where $C$ is known as parameter of reference metric which
is a positive constant. In other words, $f_{\mu \nu }$ only
depends on the spatial components $h_{ij}$ of the spacetime metric
(\ref{Metric}). Using the mentioned information and ansatz, we can extract
the explicit functional forms of $U_{i}$'s in the following forms
\begin{eqnarray}
\mathcal{U}_{1} &=&\frac{2C}{r},~~~~\;\mathcal{U}_{2}=\frac{2C^{2}}{r^{2}},
\notag \\
\mathcal{U}_{i} &=&0,~~~~~i>2.
\end{eqnarray}

Considering the field equation (\ref{field1}), the static spherical
metric (\ref{Metric}), and the mentioned reference metric (\ref{referenceM}%
), we can obtain the metric function $S(r)$, in the following form
\cite{TOV massive}%
\begin{equation}
S(r)=1-m^{2}C\left( \frac{c_{1}r}{2}+c_{2}C\right) -\frac{2GM(r)}{c^{2}r},
\label{4g(r)}
\end{equation}%
in which $M(r)=\int 4\pi r^{2}\rho (r)dr$. After some calculations,
we can extract the modified TOV equation in Vegh's massive gravity as \cite%
{TOV massive}%
\begin{equation}
\frac{dP}{dr}=\frac{G(c^{2}M(r)+4\pi r^{3}P)-\frac{m^{2}r^{2}c_{1}c^{4}C}{4}%
}{\left( \frac{m^{2}c_{1}c^{2}r^{2}C}{2}+2GM(r)+c^{2}r\left(
m^{2}c_{2}C^{2}-1\right) \right) c^{2}r}\left( c^{2}\rho +P\right) .
\label{TOV}
\end{equation}

Considering the obtained modified TOV equation in Vegh's massive
gravity, we want to investigate the properties of white dwarfs in this
theory of gravity in the next sections.

Before we continue our study about white dwarfs in Vegh's massive
gravity, we want to give a brief dimensional analysis of the parameters of
massive and reference metric. It is notable that all terms of the metric
function must be dimensionless, i.e. $\frac{m^{2}c_{1}C}{2}r$\textbf{, }$%
m^{2}c_{2}C^{2}$, and $\frac{2GM(r)}{c^{2}r}$
are dimensionless. Also, in dimensional analysis we know that
$[M(r)]=[m]=M$  (Mass), and $[r]=L$ (Length).
So, the dimensional
interpretation of massive terms are%
\begin{equation*}
\lbrack C]=L,~~~~\&~~~~[c_{i}]=M^{-2}L^{-2},~~~~i=1,2.
\end{equation*}

According to this fact that the action of massive gravity (\ref%
{Action}) is dimensionless, we find that dimensional interpretations of $%
\mathcal{R}$ and all of terms $m^{2}\sum_{i}^{4}c_{i}\mathcal{U}%
_{i}(g,f)$ are $L^{-2}$. Remembering that $%
[m^{2}c_{i}]=L^{-2}$, one can conclude that $U_{i}$'s in
Eq. (\ref{Action}), are dimensionless. As we know, the dimension of the
cosmological constant ($\Lambda $) is $L^{-2}$, so $%
m^{2}c_{i}$ terms may play the role of the pressure in the extended
phase space (see Refs. \cite{Kubiznak}, for more details).

\section{Equation of state}

We use the Chandrasekhar's equation of state (EoS), which are constituted
from electron degenerate matter,
\begin{equation}
k_{F}=\hbar (3\pi ^{2}\rho /(m_{p}\mu _{e}))^{1/3}
\end{equation}%
and
\begin{equation}
P=\frac{8\pi c}{3(2\pi \hbar )^{3}}\int_{0}^{k_{F}}\frac{k^{2}}{%
(k^{2}+m_{e}^{2}c^{2})^{1/2}}k^{2}dk,  \label{P}
\end{equation}%
where $k$ is the momentum of electrons. $m_{p}$ is the mass of proton. $\mu
_{e}$ is the mean molecular weight per electron (we choose $\mu _{e}=2$ for
our work). $\hbar =h/{2\pi }$, where $h$ is the Plank's constant. The
Chandrasekhar's EoS of the electron degenerate matter was shown in Fig. \ref%
{eos}.

\begin{figure}[tbp]
\centering
\includegraphics [width=0.6\textwidth]{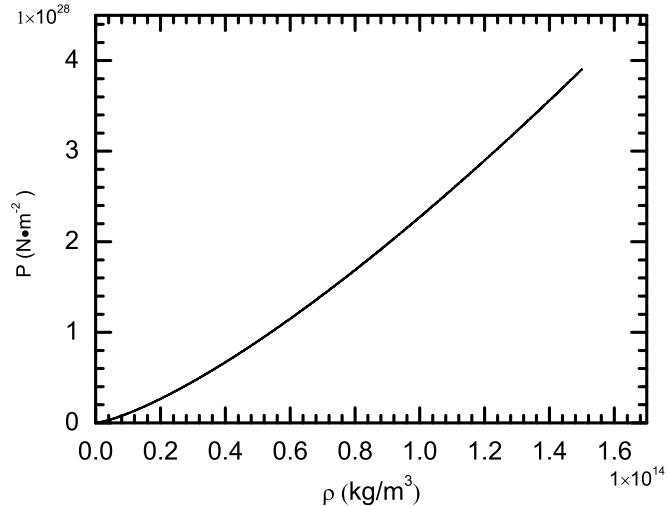}
\par
\caption{Chandrasekhar's equation of state.}
\label{eos}
\end{figure}
The Chandrasekhar's EoS is one of famous EoSs for studying the structure of
white dwarfs. In this regards, we review some properties of this EoS such
as; energy conditions, stability and Le Chatelier's principle.

The Chandrasekhar's EoS satisfies the energy conditions such as the null
energy condition (NEC), weak energy condition (WEC), strong energy condition
(SEC) and dominant energy condition (DEC) at the center of white dwarfs.
These conditions are as%
\begin{eqnarray}
NEC &\rightarrow &\ P_{c}+\rho _{c}\geq 0,  \label{11} \\
WEC &\rightarrow &\ P_{c}+\rho _{c}\geq 0,\ \ \ \&\ \ \rho _{c}\geq 0,
\label{22} \\
SEC &\rightarrow &\ P_{c}+\rho _{c}\geq 0,\ \ \ \&\ \ 3P_{c}+\rho _{c}\geq 0,
\label{33} \\
DEC &\rightarrow &\ \rho _{c}>\left\vert P_{c}\right\vert ,  \label{44}
\end{eqnarray}%
where $P_{c}$ and $\rho _{c}$\ are the pressure and density at the center of
white dwarfs ($r=0$), respectively. Using Fig. \ref{eos} and the mentioned
conditions (\ref{11}-\ref{44}), our results are presented in table \ref%
{tab11}. According to Fig. \ref{eos} and table \ref{tab11}, we observe that
all energy conditions are satisfied.
\begin{table}[tbp]
\caption{Energy conditions at the center of obtained white dwarfs of
Chandrasekhar's EoS.}
\label{tab11}
\begin{center}
\begin{tabular}{||c|c|c|c|c|c||}
\hline\hline
$\rho_{c} (10^{12}\frac{kg}{m^3})$ & $P_{c} (10^{12}\frac{kg}{m^3})$ & $NEC$
& $WEC$ & $SEC$ & $DEC$ \\ \hline\hline
$150.6590$ & $4.0515$ & $\checkmark$ & $\checkmark$ & $\checkmark$ & $%
\checkmark$ \\ \hline
\end{tabular}%
\end{center}
\end{table}

\subsection{Stability}

\begin{figure}[tbp]
\centering
\includegraphics [width=0.5\textwidth]{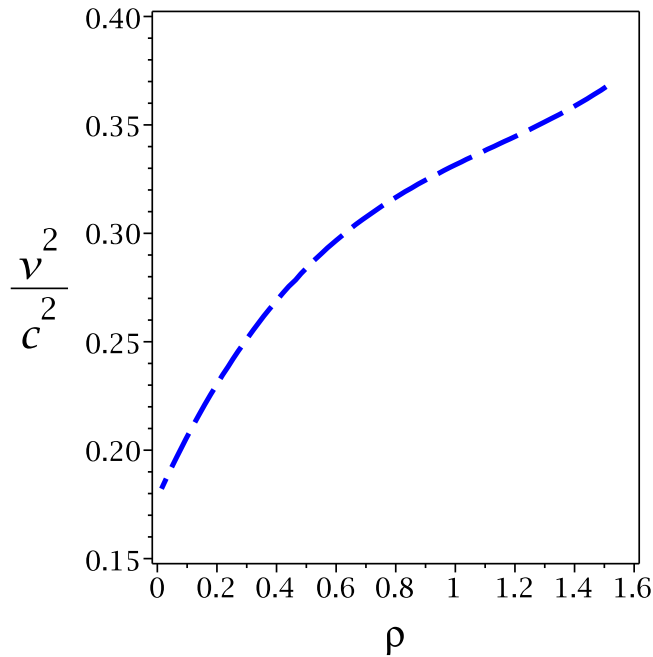}
\caption{Sound speed ($v^{2}/c^{2}\times 10^{-18}$) versus density ($\protect%
\rho \times 10^{14}$ (kg/$m^{3}$)).}
\label{Fig2}
\end{figure}

In order to evaluate the Chandrasekhar's EoS for a physically acceptable
model, one expects that the velocity of sound ($v=\sqrt{\frac{dP}{d\rho }}$)
be less than the light's velocity ($c$) \cite{Herrera1992,AbreuHN2007}. In
other words, stability condition is in the form $0\leq v^{2}\leq c^{2}$.
Therefore, by considering this stability condition and Fig. \ref{eos}, and
comparing them with diagrams related to speed of sound-density relationship
in Fig. \ref{Fig2}, it is evident that this EoS satisfies the inequality $%
0\leq v^{2}\leq c^{2}$.


\subsection{Le Chatelier's principle}

There is another important principle which is related to the matter of star
and called Le Chatelier's principle. The matter of star satisfies $dP/d\rho
\geq 0$, which is a necessary condition of a stable body both as a whole and
also with respect to the non-equilibrium elementary regions with spontaneous
contraction or expansion (Le Chatelier's principle), see Ref. \cite%
{Glendenning}, for more details. Our calculation show that, Le Chatelier's
principle is established for the Chandrasekhar's EoS (see Fig. \ref{Fig2}).

Our investigations indicated that the Chandrasekhar's EoS satisfies both
energy and stability conditions, and also this EoS admits Le Chatelier's
principle. Therefore, this EoS is a suitable EoS. It is notable that, there
are some realistic EoSs in order to have a good view of the behavior of
white dwarfs. However, in this work we consider the Chandrasekhar's EoS,
because we want to focus on the effects of modified gravity on the
structure of white dwarfs.

\section{Properties of white dwarfs}

Using the famous Chandrasekhar limit, the mass limit of white
dwarf is obtained in Newtonian, Special Relativity, GR and
Non-relativistic Newtonian theories\ in the range
$1.41-1.45M_{\odot }$ \cite{Carvalho}. On the other hand, the
explosion of peculiar SN Ia provokes us to rethink the maximum
mass of white dwarfs. Hence, the maximum mass of white dwarf is
still an open question. Here, we would like to see whether the
maximum mass of white dwarf in massive gravity and by employing
the Chandrasekhar's EoS can be more than this limit ($1.45M_{\odot
}$). Then we want to study the effects of massive's parameter on
properties of the white dwarfs such as; Schwarzschild radius, the
Kretschmann scalar and dynamical stability. It is notable
that in this paper, we consider the graviton mass as
$10^{-32}eV/c^{2}=1.78\times 10^{-65}~g$, which was
extracted by A. F. Ali and S. Das in Ref. \cite{AliD}. They have
shown from theoretical considerations, that if the graviton has mass, its value will be about $%
10^{-32}eV/c^{2}$, or $1.78\times 10^{-65}~g$. This
estimate is consistent with those obtained from experiments, including the
recent gravitational wave detection in the advanced LIGO/Virgo. Our results
indicate that by considering the special values for the parameters of
massive gravity, the maximum mass of white dwarf is an increasing
(decreasing) function of $C$ \ ($m^{2}c_{2}$), see tables \ref{tab12} and %
\ref{tab13}. Our calculations show that the maximum mass of white dwarf in
massive gravity can be more than Chandrasekhar limit ($M_{Max}>1.45M_{\odot
} $). In other words, our results predict that the mass of white dwarfs in
this gravity can be in the range upper than $2M_{\odot }$. Also, considering
the values of $m^{2}c_{2}\geq -10^{-3}$ and $C\leq 10^{-2}$, the maximum
mass and radius of white dwarfs reduce to the obtained results of GR. It is
notable that the variation of $m^{2}c_{1}$ has very interesting effect. In
this case, by variation $m^{2}c_{1}$, the maximum mass and radius of white
dwarfs are constant (see the table \ref{tabl4}).
\begin{table}[tbp]
\caption{Structure properties of white dwarf in massive gravity for $C=1$
and $m^{2}c_{1}=1\times 10^{-11}$.}
\label{tab12}\centering
\begin{tabular}{||c|c|c|c|c|c||}
\hline\hline
$m^2c_{2}$ & $M_{max}(M_\odot)$ & $R(\mathrm{km)}$ & $R_{Sch}(\mathrm{km)}$
& $\bar{\rho}(10^{12}\mathrm{kg~ m^{-3})}$ & $K ({m^{-4})}$ \\ \hline\hline
$-1\times10^{-4}$ & 1.41 & 871 & 4.17 & 1.01 & $6.37\times10^{-32}$ \\ \hline
$-1\times10^{-3}$ & 1.41 & 871 & 4.17 & 1.02 & $6.89\times10^{-30}$ \\ \hline
$-1\times10^{-2}$ & 1.43 & 875 & 4.19 & 1.02 & $6.82\times10^{-28}$ \\ \hline
$-1\times10^{-1}$ & 1.63 & 913 & 4.37 & 1.02 & $5.76\times10^{-26}$ \\ \hline
$-2\times10^{-1}$ & 1.86 & 954 & 4.57 & 1.02 & $1.93\times10^{-25}$ \\ \hline
$-4\times10^{-1}$ & 2.34 & 1030 & 4.93 & 1.02 & $5.68\times10^{-25}$ \\
\hline
$-6\times10^{-1}$ & 2.86 & 1101 & 5.27 & 1.02 & $9.80\times10^{-25}$ \\
\hline
$-8\times10^{-1}$ & 3.41 & 1168 & 4.79 & 1.02 & $1.38\times10^{-24}$ \\
\hline\hline
\end{tabular}%
\end{table}

\begin{table}[tbp]
\caption{Structure properties of white dwarf in massive gravity for $%
m^{2}c_{1}=10^{-11}$ and $m^{2}c_{2}=-2\times 10^{-1}$.}
\label{tab13}\centering
\begin{tabular}{||c|c|c|c|c|c||}
\hline\hline
$C$ & $M_{max}(M_\odot)$ & $R(\mathrm{km)}$ & $R_{Sch}(\mathrm{km)}$ & $\bar{%
\rho}(10^{12}\mathrm{kg~ m^{-3})}$ & $K ({m^{-4})}$ \\ \hline\hline
0.01 & 1.41 & 870 & 4.17 & 1.02 & $2.78\times10^{-33}$ \\ \hline
0.1 & 1.42 & 871 & 4.17 & 1.02 & $2.78\times10^{-29}$ \\ \hline
0.5 & 1.52 & 892 & 4.27 & 1.02 & $1.58\times10^{-26}$ \\ \hline
1.0 & 1.86 & 954 & 4.57 & 1.02 & $1.93\times10^{-25}$ \\ \hline
1.5 & 2.46 & 1048 & 5.02 & 1.02 & $6.71\times10^{-25}$ \\ \hline
2.0 & 3.41 & 1168 & 5.59 & 1.02 & $1.38\times10^{-24}$ \\ \hline\hline
\end{tabular}%
\end{table}

\begin{table}[tbp]
\caption{Structure properties of white dwarf in massive gravity for $C=1$
and $m^{2}c_{2}=-2\times 10^{-1}$.}
\label{tabl4}\centering
\begin{tabular}{||c|c|c|c|c|c||}
\hline\hline
$m^2c_{1}$ & $M_{max}(M_\odot)$ & $R(\mathrm{km)}$ & $R_{Sch}(\mathrm{km)}$
& $\bar{\rho}(10^{12}\mathrm{kg~ m^{-3})}$ & $K ({m^{-4})}$ \\ \hline\hline
$1\times10^{-13}$ & 1.86 & 953 & 4.56 & 1.02 & $1.94\times10^{-25}$ \\ \hline
$1\times10^{-12}$ & 1.86 & 953 & 4.56 & 1.02 & $1.94\times10^{-25}$ \\ \hline
$1\times10^{-11}$ & 1.86 & 954 & 4.57 & 1.02 & $1.93\times10^{-25}$ \\ \hline
$1\times10^{-10}$ & 1.86 & 956 & 4.58 & 1.02 & $1.91\times10^{-25}$ \\ \hline
$-1\times10^{-13}$ & 1.86 & 953 & 4.56 & 1.02 & $1.94\times10^{-25}$ \\
\hline
$-1\times10^{-12}$ & 1.86 & 953 & 4.56 & 1.02 & $1.94\times10^{-25}$ \\
\hline
$-1\times10^{-11}$ & 1.86 & 953 & 4.56 & 1.02 & $1.94\times10^{-25}$ \\
\hline
$-1\times10^{-10}$ & 1.85 & 950 & 4.55 & 1.02 & $1.96\times10^{-25}$ \\
\hline\hline
\end{tabular}%
\end{table}

\begin{figure}[tbp]
\centering
\includegraphics[width=.45\textwidth]{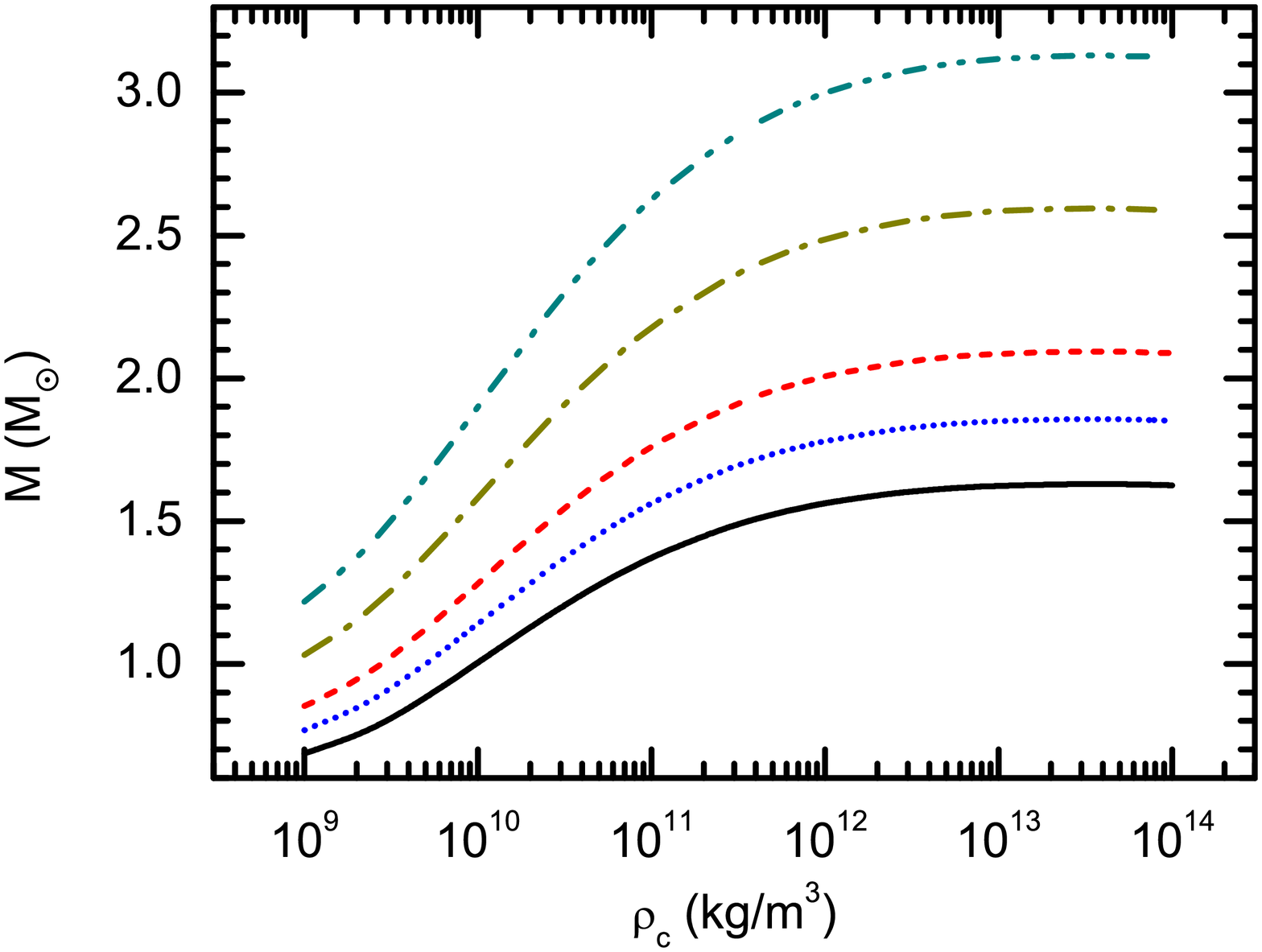} \hfill %
\includegraphics[width=.45\textwidth]{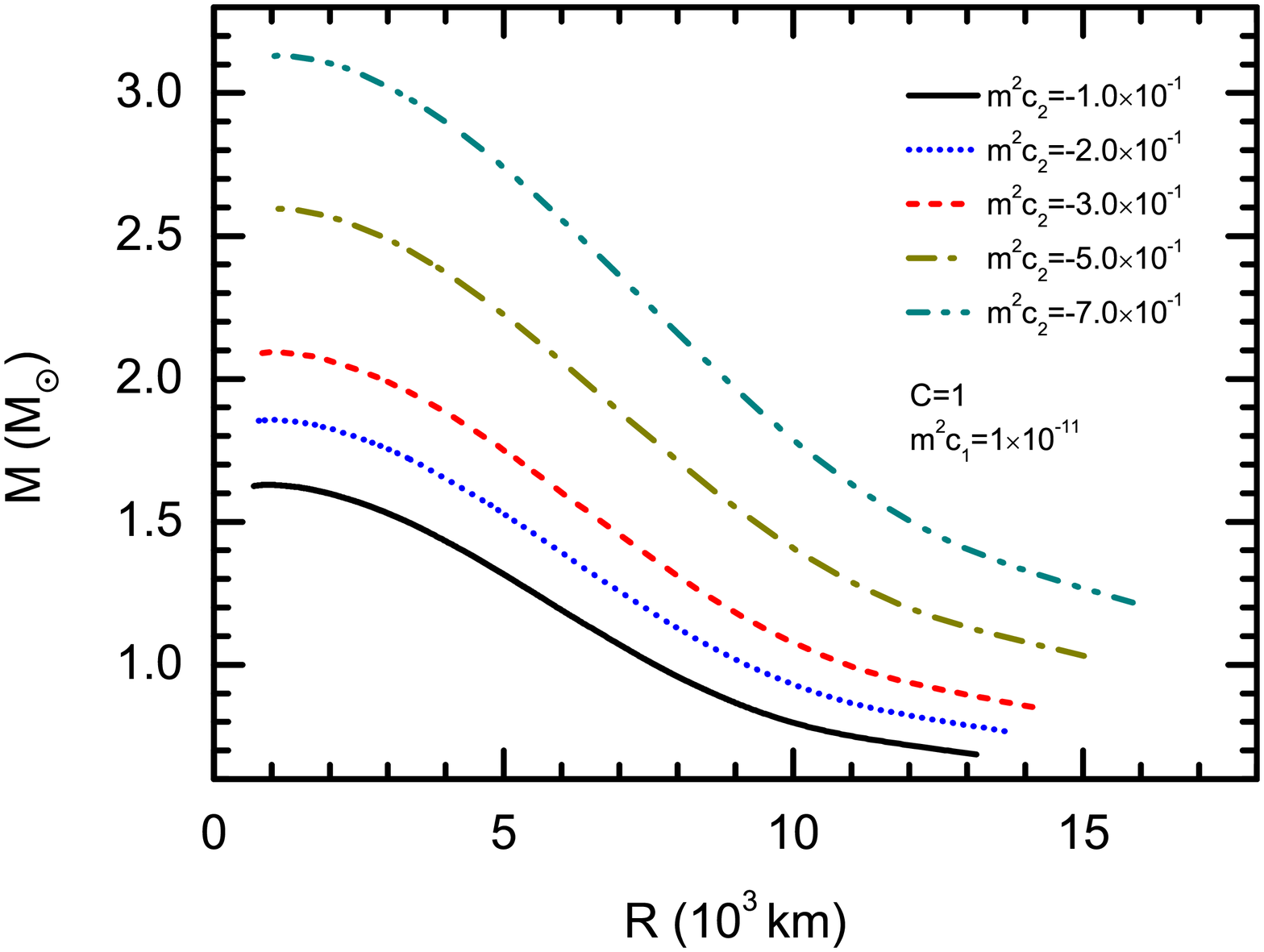}
\caption{Gravitational mass versus central density(radius) for $C=1$ and $%
m^2c_{1}=1\times10^{-11}$. Left diagrams: gravitational mass versus central
mass density for $m^2c_{2}=-1.0\times10^{-1}$ (solid line), $%
m^2c_2=-2.0\times10^{-1}$ (dotted line), $m^2c_2=-3.0\times10^{-1}$ (dashed
line), $m^2c_2=-5.0\times10^{-1}$ (dashed-dotted line) and $%
m^2c_2=-7.0\times10^{-1}$ (dashed-dotted-dotted line). Right diagrams:
gravitational mass versus radius for $m^2c_{2}=-1.0\times10^{-1}$ (solid
line), $m^2c_2=-2.0\times10^{-1}$ (dotted line), $m^2c_2=-3.0\times10^{-1}$
(dashed line), $m^2c_2=-5.0\times10^{-1}$ (dashed-dotted line) and $%
m^2c_2=-7.0\times10^{-1}$ (dashed-dotted-dotted line).}
\label{MRI}
\end{figure}

\begin{figure}[tbp]
\centering
\includegraphics[width=.45\textwidth]{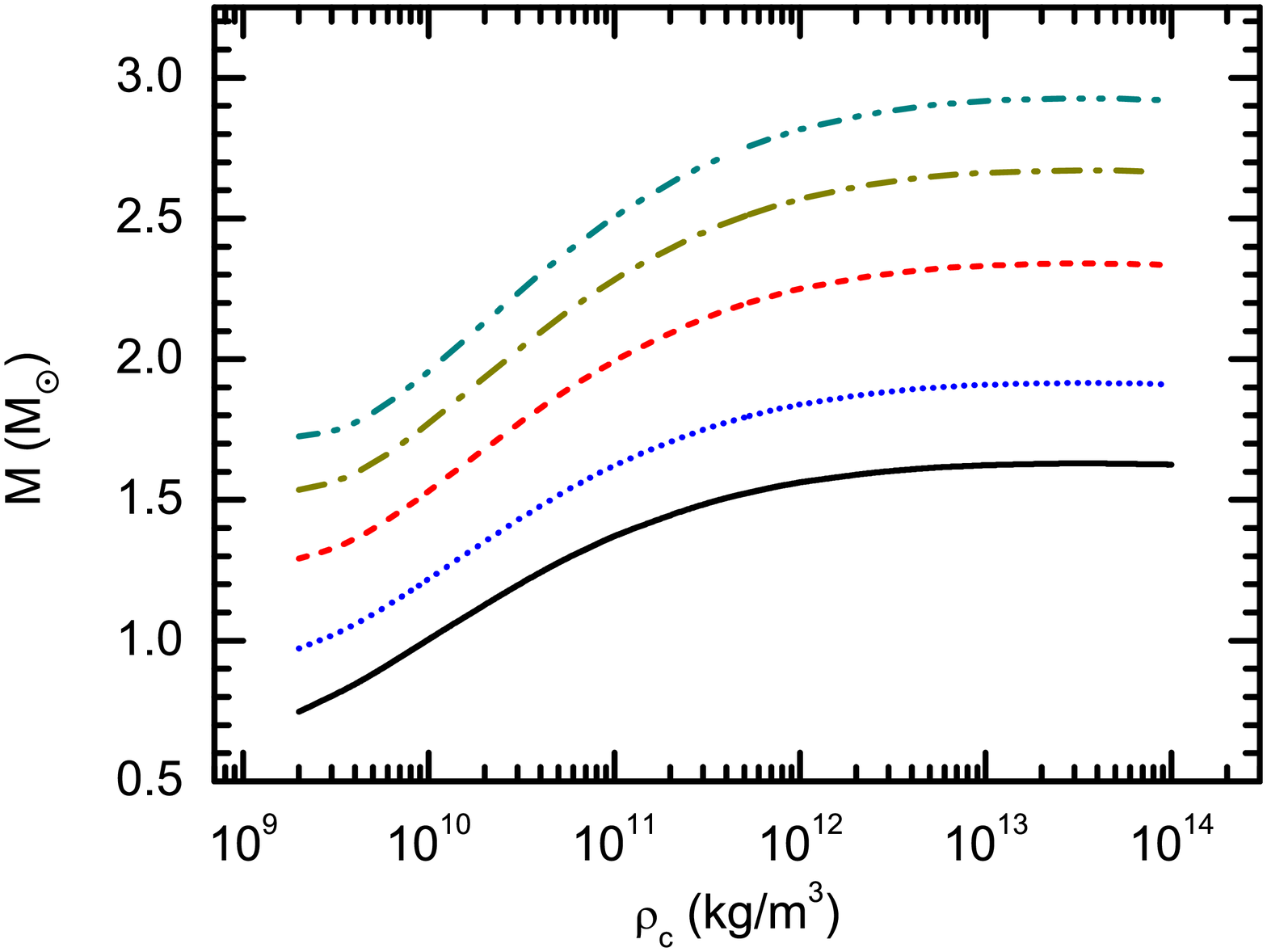} \hfill %
\includegraphics[width=.45\textwidth]{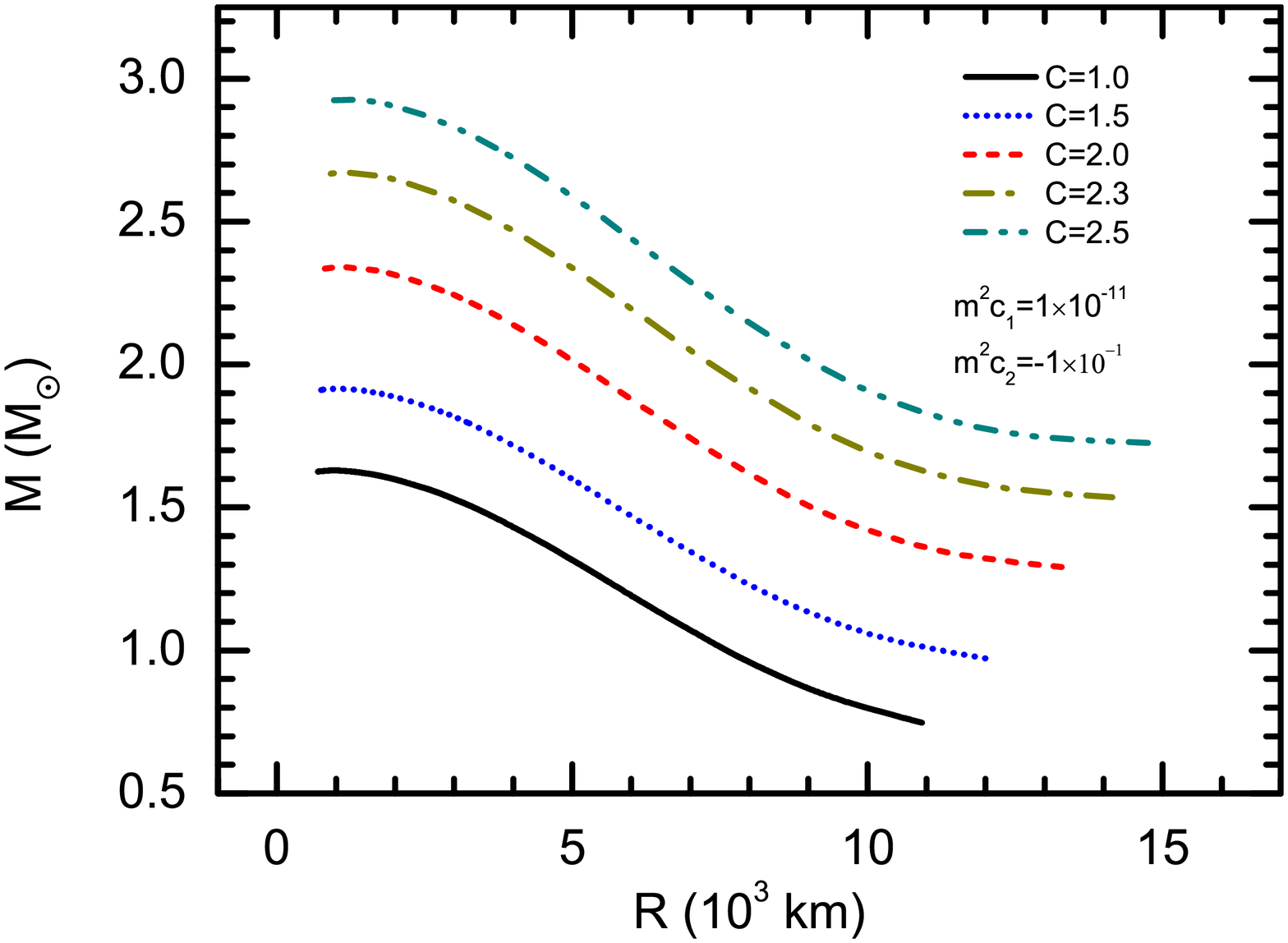}
\caption{Gravitational mass versus central density (radius) for $%
m^2c_1=1\times10^{-11}$ and $m^2c_{2}=-1\times10^{-1}$. Left diagrams:
gravitational mass versus central mass density for $C=1.0$ (solid line), $%
C=1.5$ (dotted line), $C=2.0$ (dashed line), $C=2.3$ (dashed-dotted line)
and $C=2.5$ (dashed-dotted-dotted line). Right diagrams: gravitational mass
versus radius for $C=1.0$ (solid line), $C=1.5$ (dotted line), $C=2.0$
(dashed line), $C=2.3$ (dashed-dotted line) and $C=2.5$
(dashed-dotted-dotted line).}
\label{MRII}
\end{figure}

In order to do more investigation, we plot the mass of white dwarf versus
the central mass density ($M-\rho _{c}$), for different parameters of
massive gravity and reference metric in left panels of Figs. \ref{MRI} and %
\ref{MRII}. These figures show that, the maximum mass of white dwarf
increases when $m^{2}c_{2}$ decreases or $C$ increases.
In addition, the variation of maximum mass versus radius ($M-R$) is also
shown in right panels of Figs. \ref{MRI} and \ref{MRII}.

Our calculations show that by considering fixed values for $C$ and $%
m^{2}c_{1}$, in constant radius (for example $R=5\times 10^{3}km$), by
decreasing $m^{2}c_{2}$, the mass of white dwarfs increase (see right panel
in Fig. \ref{MRI}). Also, there is the same behavior for $C$ (see right
panel in Fig. \ref{MRII}). In other words, in $R=$constant, by varying the
parameters of massive gravity and the reference metric, the mass of white
dwarfs change. This result shows that the density within of white dwarf
depends on the parameters of this theory of gravity, so that it increases
when the mass of white dwarf increases.

Here we can ask this question: why do the maximum mass of white dwarfs
increase by varying the parameters of this theory? The strength of gravity
may change by varying the parameters of massive gravity and the reference
metric. In other words, by increasing ($C$) or decreasing ($m^{2}c_{2}$)
parameters of this theory, the strength of gravity may decrease. As we know,
there is a balance between the internal pressure (which origin it is
electron degenerate) and gravitational force. Decreasing the strength of
gravity, a star can bear more mass in order to keep this balance. Therefore,
the maximum mass of white dwarf increases by increasing ($C$) or decreasing (%
$m^{2}c_{2}$).

For completeness, in the following, we investigate other properties of white
dwarf in massive gravity such as the Schwarzschild radius, average density,
Kretschmann scalar and dynamical stability.

\subsection{Modified Schwarzschild Radius}

The Schwarzschild radius for this gravity is obtained as \cite{TOV massive}
\begin{equation}
R_{Sch}=\frac{c\left( 1-m^{2}c_{2}C^{2}\right) }{m^{2}cc_{1}C}-\frac{\sqrt{%
c^{2}\left( m^{2}c_{2}C^{2}-1\right) ^{2}-4m^{2}c_{1}CGM}}{m^{2}cc_{1}C}.
\label{Sch}
\end{equation}

The Schwarzschild radius of white dwarfs are obtained in tables \ref{tab12}
and \ref{tab13}. These results show that by increasing the maximum mass and
radius of white dwarfs, the Schwarzschild radius increases and the obtained
white dwarfs in massive gravity with mass more than the Chandrasekhar limit
are out of the Schwarzschild radius (see tables \ref{tab12} and \ref{tab13}%
). In other words, different parameters of massive gravity and the reference
metric have different behavior on the Schwarzschild radius. For example, by
considering the negative value of $m^{2}c_{2}$ and increasing it, the
Schwarzschild radius increases (see table \ref{tab12}). Also, by increasing $%
C$, the Schwarzschild radius increases (see table \ref{tab13}). On the other
hand, by considering the positive (negative) values of $m^{2}c_{1}$ and
increasing (decreasing) $m^{2}c_{1}$, the Schwarzschild radius does not
change (see table \ref{tabl4}).

\subsection{Average Density}

We can evaluate the average density by using the obtained maximum mass and
radius of white dwarfs in the massive gravity from the perspective of a
distant observer (or a observer outside the neutron star). So the average
density of white dwarf is given
\begin{equation}
\overline{\rho }=\frac{3M}{4\pi R^{3}},  \label{density}
\end{equation}%
where the results for variation of the massive parameters and the reference
metric are presented in the tables \ref{tab12}, \ref{tab13} and \ref{tabl4}.
There is an interesting result about the average density of white dwarf in
massive gravity from the perspective of a observer outside the white dwarf.
By variations of the different parameters, the average density remains fixed
(see tables \ref{tab12}, \ref{tab13} and \ref{tabl4}).

In order to investigate the strength of gravity, we study the Kretschmann
scalar in the presence of nonzero graviton mass in the following subsection.

\subsection{Kretschmann scalar}

Spacetime curvature is a quantity that shows the strength of
gravity. It is worth mentioning that in the Schwarzschild
spacetime, the components of Ricci scalar ($R$) and the Ricci
tensor ($R_{\mu \nu }$) are zero outside the star, and these
quantities do not give us any information about the spacetime
curvature (or the strength of gravity). In order to investigate
the curvature of spacetime in more details, we use another
quantity. The quantity is the Riemann tensor ($R_{\mu \nu \gamma
\delta }$). It is notable that\textbf{, }the Riemann tensor may
have more components, and, for simplicity, we can study the
Kretschmann scalar for measurement of the curvature in a vacuum.
After some calculations, we can obtain the curvature at
the surface of a white dwarf in massive gravity as
\begin{equation}
K=R_{\mu \nu \gamma \delta }R^{\mu \nu \gamma \delta }=\frac{%
2m^{4}c_{1}^{2}C^{2}}{R^{2}}+\frac{4m^{4}c_{1}c_{2}C^{3}}{R^{3}}+\frac{%
4m^{4}c_{2}^{2}C^{4}}{R^{4}}+\frac{16m^{2}c_{2}C^{2}GM}{c^{2}R^{5}}+\frac{%
48G^{2}M^{2}}{c^{4}R^{6}},
\end{equation}%
where in the absence of graviton mass ($m=0$), the above
equation reduces to the Kretschmann scalar in Einstein gravity as
$K=\frac{ 48G^{2}M^{2}}{c^{4}R^{6}}$,
\cite{Psaltis,Eksi,EslamPanah}. Our results show that by
considering nonzero graviton mass, the strength of gravity from
the perspective of a distant observer increases when the mass of
white dwarfs increases (see tables \ref{tab12} and \ref{tab13}).

\subsection{Dynamical Stability}

Another important quantity is related to the dynamical stability of white
dwarfs in massive gravity. Chandrasekhar introduced the dynamical stability
of stellar model against the infinitesimal radial adiabatic perturbation in
ref. \cite{Chandrasekhar}. Some authors applied it to astrophysical cases
\cite{BardeenTM,Kuntsem,Mak,Kalam}. The adiabatic index ($\gamma $) is
defined in the following form

\begin{equation}
\gamma =\frac{\rho c^{2}+P}{c^{2}P}\frac{dP}{d\rho }.
\end{equation}

\begin{figure}[tbp]
\centering
\includegraphics[width=.45\textwidth]{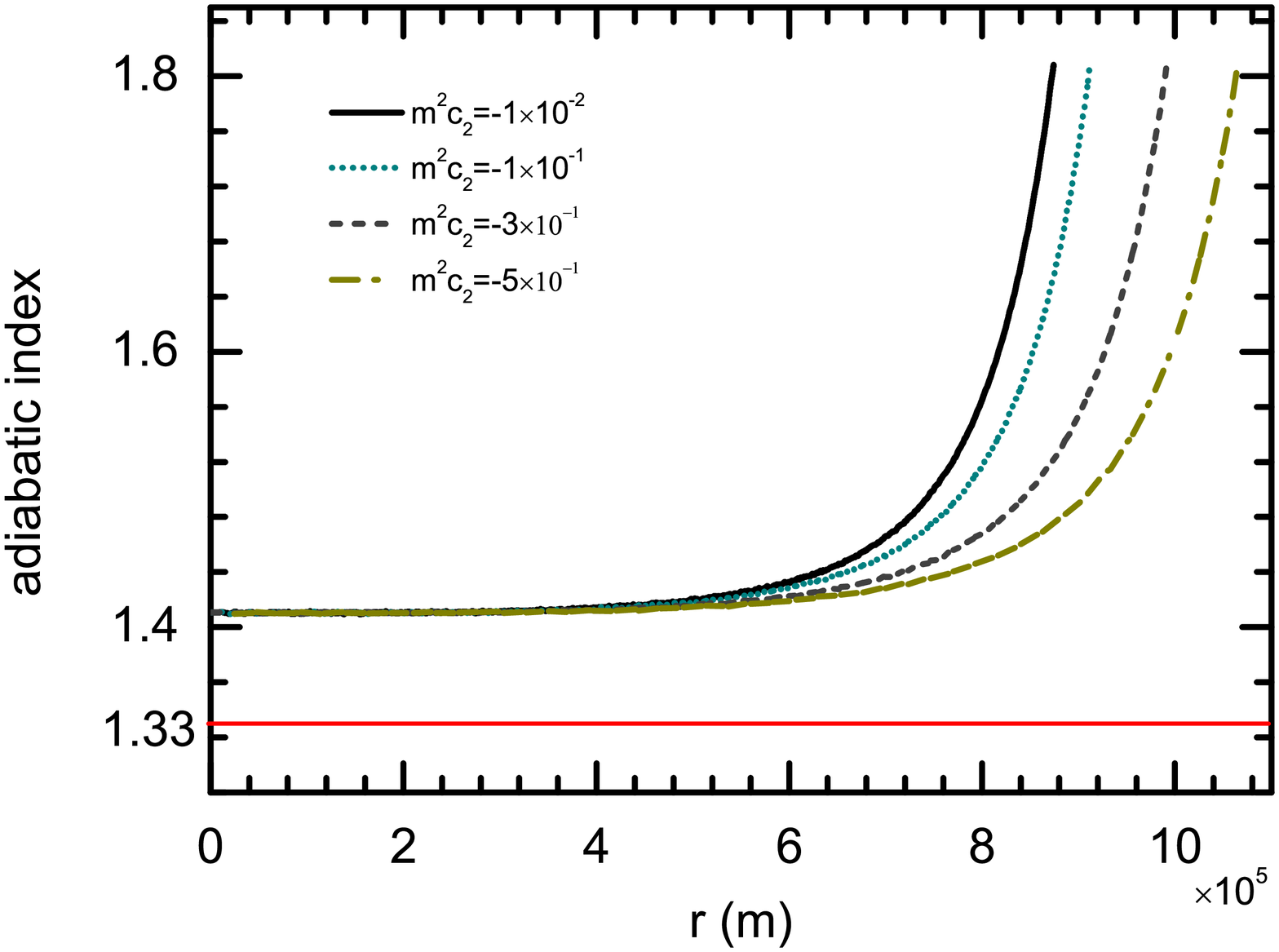} \hfill %
\includegraphics[width=.45\textwidth]{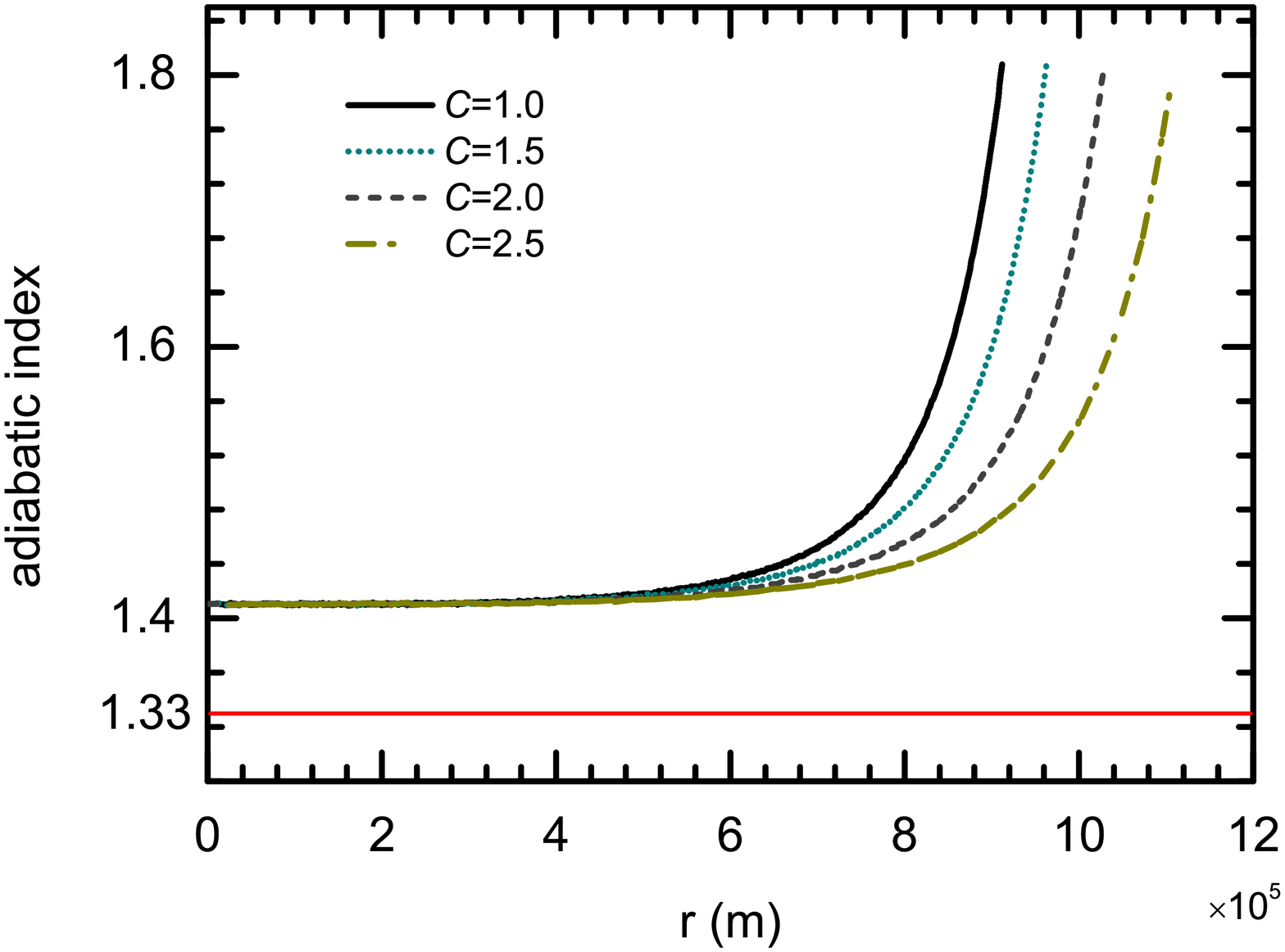}
\caption{Adiabatic index versus radius for $m^{2}c_{1}=1\times 10^{-11}$.
Left diagrams: for $C=1$, $m^{2}c_{2}=-1\times 10^{-2}$ (continuous line), $%
m^{2}c_{2}=-1\times 10^{-1}$ (dotted line), $m^{2}c_{2}=-3\times 10^{-1}$
(dashed line), $m^{2}c_{2}=-5\times 10^{-1}$ (dashed-dotted line). Right
diagrams: for $m^{2}c_{2}=-1\times 10^{-1}$, $C=1.0$ (continuous line), $%
C=1.5$ (dotted line), $C=2.0$ (dashed line), $C=2.5$ (dashed-dotted line).}
\label{gamma}
\end{figure}

We will encounter with the dynamical stability when $\gamma $ is more than $%
4/3$ ($\gamma >4/3\simeq1.33$), everywhere within the obtained
white dwarfs. So, we plot two diagrams related to $\gamma $ versus
radius for different values of $m^{2}c_{2}$ and $C$ in Fig.
\ref{gamma}. Our results show that, the super-Chandrasekhar white
dwarfs or massive white dwarfs are stable against the radial
adiabatic infinitesimal perturbations.

\begin{figure}[tbp]
\centering
\includegraphics[width=.45\textwidth]{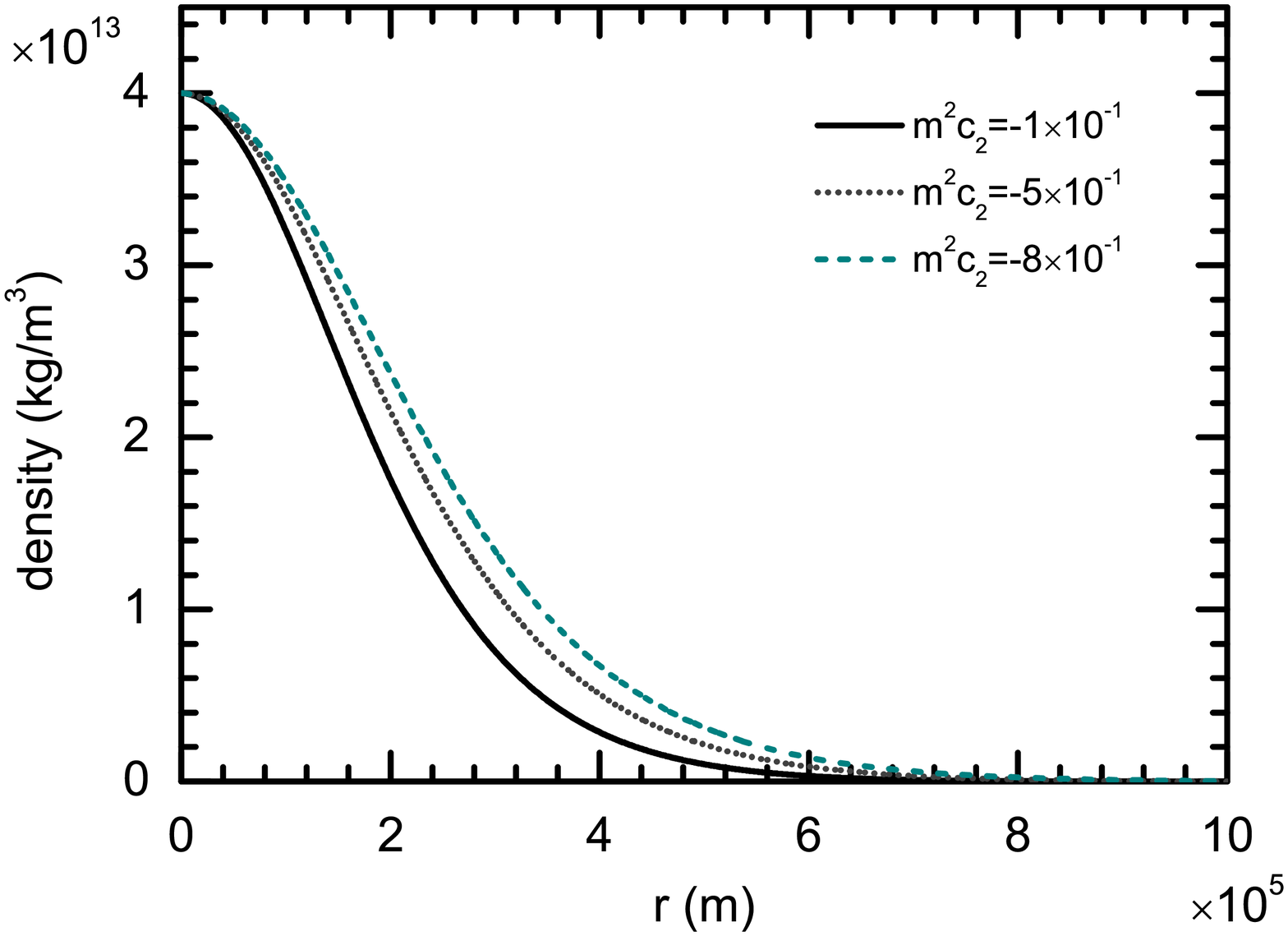} \hfill %
\includegraphics[width=.45\textwidth]{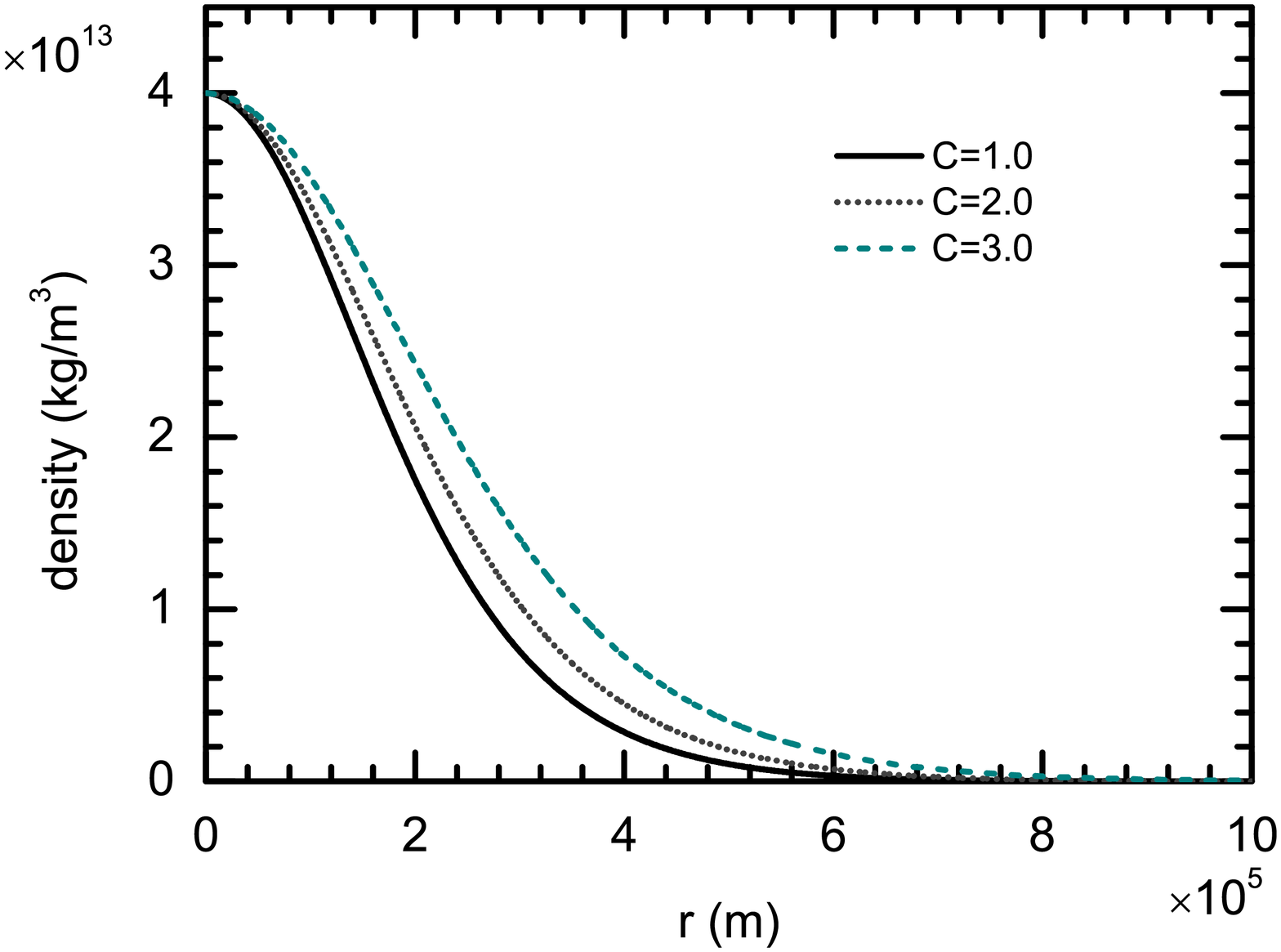}
\caption{Density versus radius for $m^{2}c_{1}=1\times 10^{-11}$. Left
diagrams: for $C=1$, $m^{2}c_{1}=-1\times 10^{-1}$ (continuous line), $%
m^{2}c_{1}=-5\times 10^{-1}$(dotted line), $m^{2}c_{1}=-8\times 10^{-1}$
(dashed line). Right diagrams: for $m^{2}c_{2}=-1\times 10^{-1}$, $C=1.0$
(continuous line), $C=2.0$ (dotted line), $C=3.0$ (dashed line).}
\label{denr}
\end{figure}

\begin{figure}[tbp]
\centering
\includegraphics[width=.45\textwidth]{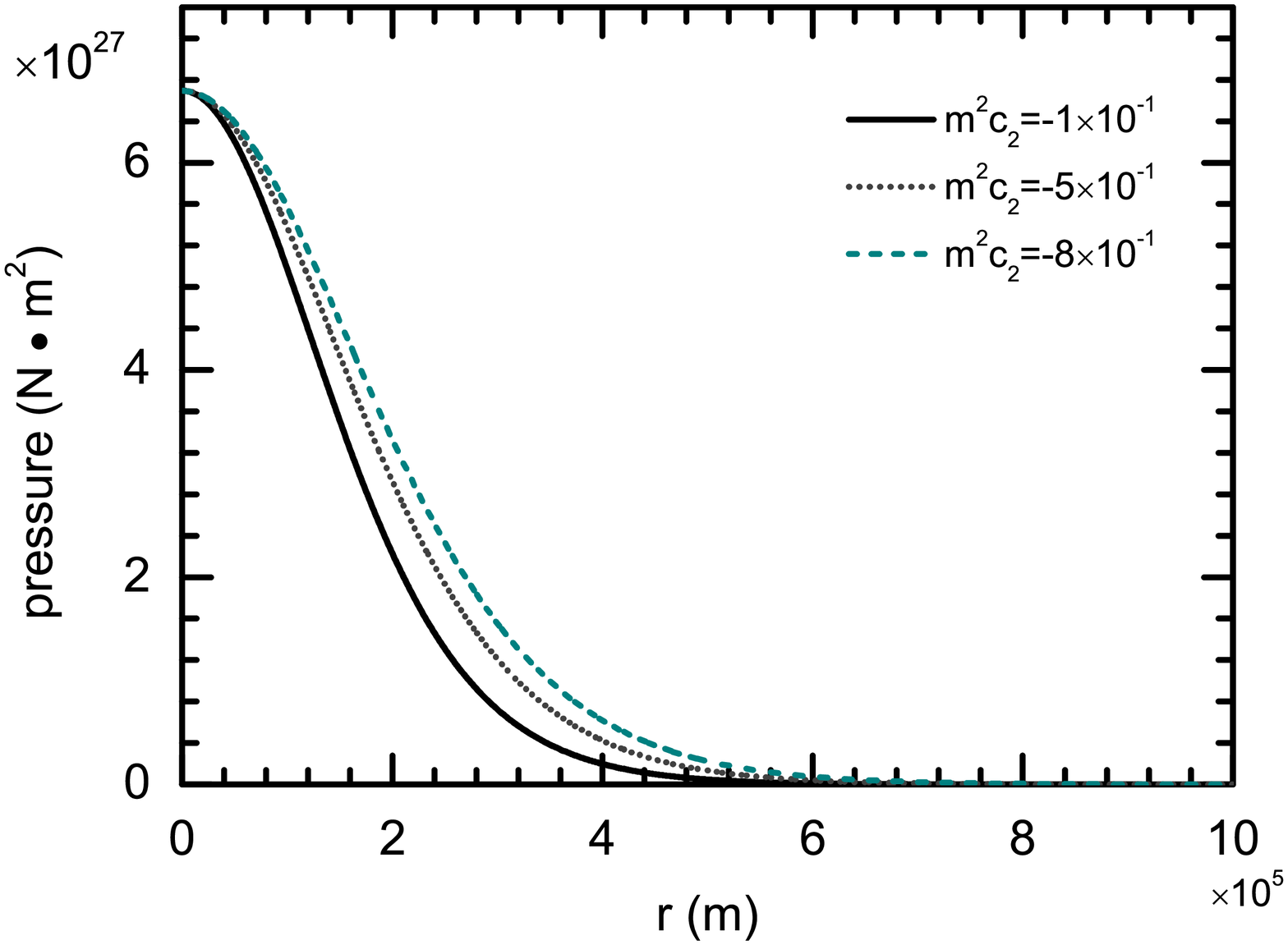} \hfill %
\includegraphics[width=.45\textwidth]{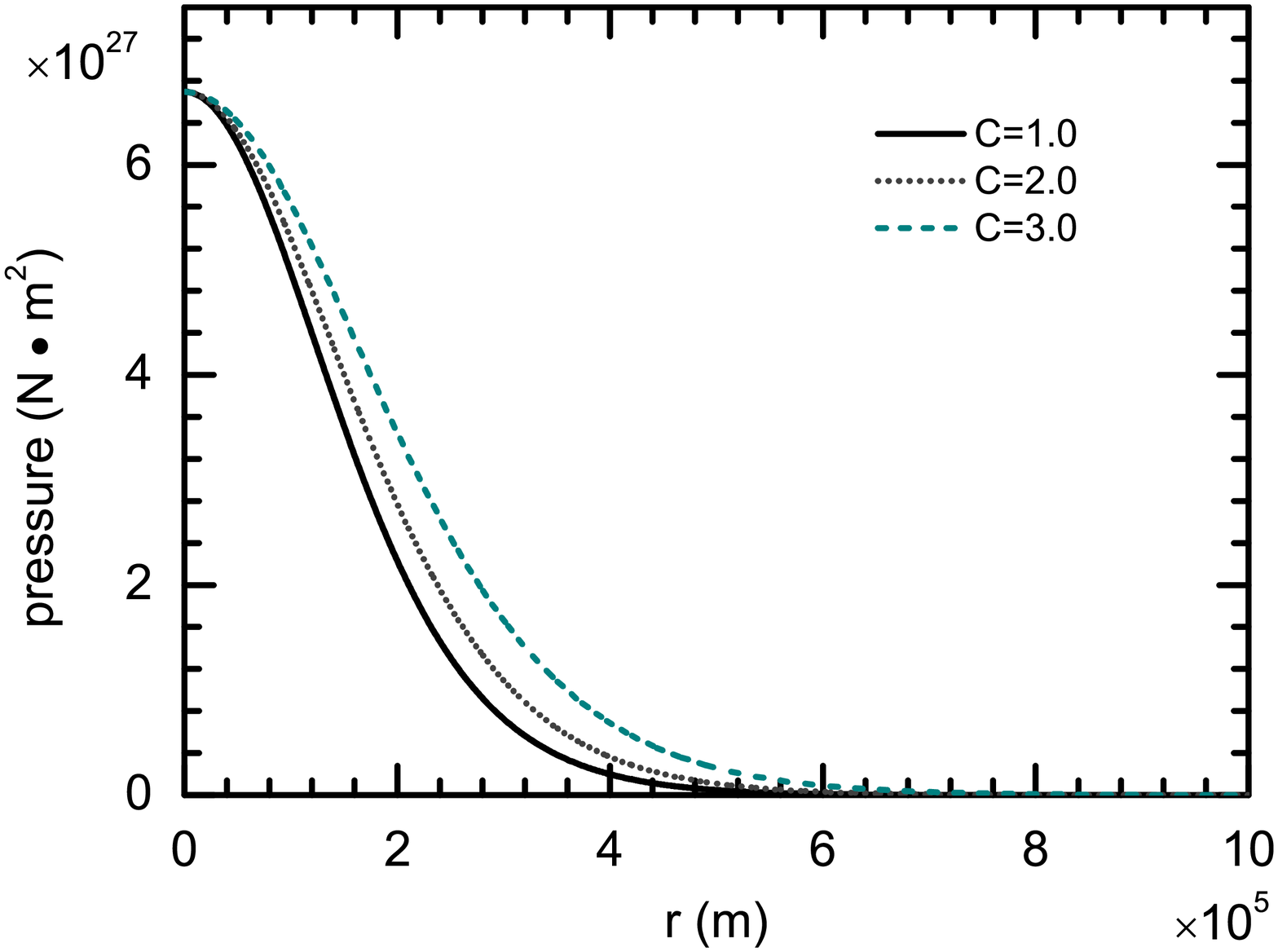}
\caption{Pressure versus radius for $m^{2}c_{1}=1\times 10^{-11}$. Left
diagrams: for $C=1$, $m^{2}c_{1}=-1\times 10^{-1}$ (continuous line), $%
m^{2}c_{1}=-5\times 10^{-1}$ (dotted line), $m^{2}c_{1}=-8\times 10^{-1}$
(dashed line). Right diagrams: for $m^{2}c_{2}=-1\times 10^{-1}$, $C=1.0$
(continuous line), $C=2.0$ (dotted line), $C=3.0$ (dashed line).}
\label{pr}
\end{figure}

Here, we want to evaluate the behavior of density and pressure versus
distance from the center to the surface of white dwarfs. For this goal, we
plot them in Figs. \ref{denr} and \ref{pr}. As one can see, the density and
the pressure are maximum at the center and they decrease monotonically
towards the boundary. These figures show that, in constant radius (for
example $r=4\times 10^{3}km$), by increasing $C$ (or decreasing $m^{2}c_{2}$%
), the density and the pressure increase.


\section{Comparison between neutron stars and white dwarfs in massive gravity%
}

In this section we want to compare the obtained neutron stars in ref. \cite%
{TOV massive} with the obtained white dwarfs in this paper.

\textbf{Radius:} the obtained radius of massive neutron stars were about $10$
km ($R_{NS}\simeq 10km$) \cite{TOV massive}, whereas the super-Chandrasekhar
white dwarfs are about $1000$ km, so the radius of these super-Chandrasekhar
white dwarfs are one hundred times larger than radius of massive neutron
stars ($R_{WD}\simeq 100R_{NS}$).

\textbf{Average density:} the average density of massive neutron stars in
massive gravity is $\overline{\rho }_{NS}\simeq 10^{15}g~cm^{-3}$ \cite{TOV
massive}, while the average density of the super-Chandrasekhar white dwarfs
in this gravity is $\overline{\rho }_{WD}\simeq 10^{9}g~cm^{-3}$. Therefore,
the average density of massive white dwarfs is less than the average density
of massive neutron stars, $\overline{\rho }_{WD}<\overline{\rho }_{NS}$ ($%
\overline{\rho }_{NS}\simeq 10^{6}\overline{\rho }_{WD}$), as we expected.

\textbf{Kretschmann scalar:} in order to investigate the strength of gravity
between massive neutron stars and massive white dwarfs, we compare the
values of their Kretschmann scalars. Our calculations of the Kretschmann
scalar for massive neutron stars are $K_{NS}\simeq 5.7\times 10^{-16}m^{-4}$%
, while for massive white dwarfs are $K_{WD}\simeq 6.7\times 10^{-25}m^{-4}$%
. Comparing these values show that, the strength of gravity of massive white
dwarfs are less than the massive neutron stars, $K_{WD}<K_{NS}$ ($%
K_{NS}\simeq 10^{9}K_{WD}$).

\textbf{Dynamical stability: }our results show that, both of them (the
massive neutron stars and the super-Chandrasekhar white dwarfs) satisfy this
condition. In other words, both of them are stable against the radial
adiabatic infinitesimal perturbations.

\section{Closing Remarks}

As we mentioned before, in order to find the massive or super-Chandrasekhar
white dwarfs, we can consider two approaches; i) improve the EoS, and ii)
modified gravity. According to this fact that GR had some problems,
and also the probability of existence massive graviton based on\ the recent
observations by the advanced LIGO/Virgo \cite{Abbott,AbbottII} collaboration%
, which had put a tight bound on graviton's mass, and another theoretical
and empirical limits on the mass of gravitons, in this work we
considered the dRGT like massive gravity which is known as Vegh's massive
gravity with a reference metric in the form of eq. (\ref{referenceM}).%
 Then we investigated the effects of this theory of gravity on the
structure of white dwarfs. We employed the Chandrasekhar's EoS and
considered the modified TOV equation in the presence of nonzero graviton
mass. Our results showed that the maximum mass of white dwarfs in massive
gravity can be more than Chandrasekhar limit ($M_{Max}>1.45M_{\odot }$),
because the strength of gravity may change by varying the parameters of this
gravity. Indeed, by increasing ($C$) or decreasing ($m^{2}c_{2}$) parameters
of the reference metric and massive theory of gravity, respectively, the
strength of gravity may decrease. As we know, there is a balance between the
internal pressure and gravitational force. Decreasing the strength of
gravity, a star can bear more mass in order to keep this balance. Therefore,
the maximum mass of white dwarf increased by increasing ($C$) or decreasing (%
$m^{2}c_{2}$). Then, we studied other properties of super-Chandrasekhar
white dwarfs such as Schwarzschild radius, average density and Kretschmann
scalar in the presence of nonzero graviton mass. Next, we evaluated the
dynamical stability in order to have physical super-Chandrasekhar white
dwarfs. For having super-Chandrasekhar white dwarfs, we obtained some
constraints on the parameters of massive gravity. Indeed, our results showed
that the value of $C$ has to be more than $0.1$ ($C>0.1$). The sign of $%
c_{2} $ was negative with the range $m^{2}c_{2}<-1\times 10^{-2}$. In
addition, the sign of $c_{1} $ could be positive or negative, so that
different values of this parameter did not affect the structure of white
dwarfs (see table \ref{tabl4}).

Recently, in Ref. \cite{conformal}, was shown that there is a
correspondence between the spherical black hole solutions in this theory of
massive gravity with conformal gravity, when $m^{2}c_{1}C$ and $%
m^{2}c_{2}C^{2}$ are in the ranges; (i) $m^{2}c_{1}C>0$
and $-2<m^{2}c_{2}C^{2}<0$, (ii) $m^{2}c_{1}C<0$ and $%
m^{2}c_{2}C^{2}<-2$, or (iii) $m^{2}c_{1}C<0$ and $%
m^{2}c_{2}C^{2}>0 $. The results obtained in this paper were
consistent with case (i). Also, this range (case (i)) was valid for neutron
stars in massive gravity \cite{TOV massive}.

Finally, in order to have a better view of these super-Chandrasekhar white
dwarfs in massive gravity, we compared the obtained massive neutron stars in
ref. \cite{TOV massive}, with super-Chandrasekhar white dwarfs in the last
section.

Briefly, we obtained the quite interesting results from massive gravity for
the white dwarfs such as;

I) Prediction of maximum mass for white dwarfs more than the Chandrasekhar
limit ($M>1.45M_{\odot }$), due to the existence of nonzero graviton mass.
In other words, super-Chandrasekhar white dwarfs in massive gravity were
acceptable.

II) The super-Chandrasekhar white dwarfs in massive gravity are dynamically
stable.

III) Considering different values of parameters of massive gravity, the
strength of gravity from the perspective of a distant observer by increasing
the mass of white dwarf increased.

IV) Density inside white dwarfs increased due to increasing the mass of
white dwarf, which was one of the effects of massive gravity.

V) Super-Chandrasekhar white dwarfs imposed some constraints on parameters
of massive theory.

Finally, it is notable that rotating, slowly rotating and magnetized white
dwarfs \cite{rootI,rootIII,rootIV,rootV,rootVI,rootVII,rootVIII,rootIX,rootX}
in the context of massive gravity are interesting topics. In addition, it
will be very interesting if we use another realistic equation of state in
order to have a good view of the behavior of white dwarfs in massive
gravity. We leave these issues for future works.

\begin{acknowledgements}
We thank an anonymous referee for useful comments. We also
acknowledge S. Panahiyan and S. H. Hendi for reading the
manuscript. B. Eslam Panah thanks Shiraz University Research
Council. The work of B.E.P. has been supported
financially by Research Institute for Astronomy and Astrophysics
of Maragha (RIAAM) under research project No. 1/5750-49. The work of
H.L.L. has been supported financially by the National
Natural Science Foundation of China under No. 11803026.

\end{acknowledgements}

\end{document}